\documentclass[%
 reprint,
 amsmath,amssymb,
 aps,longbibliography,
 prl
]{revtex4-1}

\usepackage{amsthm}
\usepackage{lmodern}
\usepackage{graphicx}
\usepackage{dcolumn}
\usepackage{bm}
\usepackage{hyperref}
\hypersetup{linktocpage,colorlinks,citecolor={blue},pdfdisplaydoctitle=true,pdfpagemode=UseOutlines,bookmarksnumbered=true}
\usepackage{mathrsfs,dsfont}
\usepackage{color}
\usepackage{microtype}
\usepackage{isotope}

\newcommand\keff{k_\text{eff}}

\newcommand{\eg}[0]{\textit{e.g.} }

\begin{document}
\title{Arbitrarily large neutron amplification in subcritical nuclear reactors}

\author{Antoine Tilloy}
\affiliation{Max-Planck-Institut f\"ur Quantenoptik, Hans-Kopfermann-Stra{\ss}e 1, 85748 Garching, Germany}

\begin{abstract}
\noindent
In a subcritical reactor, each neutron produces only $\keff<1$ neutron per generation (asymptotically and on average), and thus the neutron population decreases exponentially in the absence of external source. The chain reaction is thus easy to stop, making such reactors inherently stable. Interestingly, and contrary to common wisdom, there is no relation between $\keff$, the external source intensity, and the output power of the reactor. In this paper, I present various possible strategies to exploit this fact, and apply them to the design of a  rudimentary multi-layer system that allows to reach an arbitrarily large number of fissions per source neutron, while keeping $\keff < 1$ fixed (I used $\keff=0.97$). The behavior of the amplifier is verified with simple Monte-Carlo transport simulations. The proposal admittedly brings complications and subtleties that need to be studied further, but, if developed, could allow to drive subcritical reactors with faint neutron sources, \eg radioisotope based.
\end{abstract}

\maketitle

\paragraph{Introduction --}
\noindent
Subcritical nuclear reactors are a serious candidate for the incineration of minor actinides produced by current light water reactors. With an asymptotic neutron multiplication $k_{\text{eff}} < 1$, they allow an intrinsically stable operation without the need for delayed neutrons, and can thus operate which much higher minor actinide content than more standard critical reactors. But neutrons have to come from somewhere, and one\footnote{There are many other obstacles to a large scale transmutation strategy of radioactive waste. The improvement of the neutron efficiency of subcritical reactors would only lift a minor hurdle.} of the drawbacks of such designs lies in the need for an efficient external neutron source. Current proposals use a spallation source driven by a high current proton accelerator. To reach an industrially meaningful actinide incineration or electricity production, the accelerator power required goes well beyond what is available off-shelf. 

The high power need comes from the requirement of a $k_{\text{eff}}$ not too close to $1$ to ensure safe operation, the choice $k_\text{eff}\simeq 0.95$ usually being made in the literature \cite{nifenecker2003accelerator}. At an intuitive level, this suggests that one neutron can produce on average $0.95$ neutrons per generation and thus a number of fissions of the order of $20$ from the first fission induced by the spallation source. While this rule of thumb may give a correct order of magnitude for standard accelerator driven subcritical reactor (ADSR) designs and explains the need for a high intensity neutron source, but it can give the wrong order of magnitude in more complicated situations.

We need to keep track of neutron generations and consider the fact that the multiplication per neutron $k_i$ can be generation dependent because of the geometry of the system. The number of neutrons produced from a single source neutron is on average:
\begin{equation}
    N = k_1 + k_1 k_2 + k_1 k_2 k_3 + \cdots = \sum_{i=1}^{+\infty} \prod_{j=1}^i k_j \overset{\text{def.}}{=} \frac{1}{1-k_s}.
\end{equation}
This $N$, or equivalently $k_s$, is what we care about when we consider the total neutron multiplication of the system, and the output power for a given source. The number $k_s$ is the equivalent multiplication we would need to get the same $N$ had the multiplication $k_i$ been generation independent. Importantly, it has nothing to do with $\keff$, which is defined as:
\begin{equation}
    \keff\overset{\text{def.}}{=} \lim_{i\rightarrow +\infty} k_i\, ,
\end{equation}
which is the asymptotic multiplication. Only $\keff$ governs the long time reactor kinematics and is thus relevant to assess criticality. We see that, in principle, there is no theoretical obstacle to a $N$ arbitrarily large ($k_s$ arbitrarily close to $1$) with a fixed subcritical $\keff$ provided the first $k_i$'s are large.

In fact, some ADSR designs obtain modest amplification from the naive $\keff=k_s$ relation with a highly enriched core near the spallation source surrounded by low enriched blankets. However, such designs typically reduce the accelerator requirements only by a rather moderate amount.

 This brings a natural question: is it possible to design a core such that $\keff$ is fixed but $N$ is large enough to substantially reduce the accelerator requirements? Can multiplication be increased to the point that a much smaller accelerator can be used or even that much weaker radioisotope sources can be used? Is there a physical limit to amplification due to neutron transport in existing materials? Is there a practical limit with current technology and engineering constraints? 
 
 \paragraph{Strategies --}

\noindent
To increase $N$ while keeping $\keff<1$ one needs to create an irreversibility or decoupling in the neutron current. Imagine we were given a perfect neutron diode, that is a material letting neutrons go in one, and only one direction. Then, we could easily stack subcritical sections, separating them by the diodes and the total would remain subcritical, with the same $\keff$. However, the neutron amplification would be enhanced geometrically with the number of layers, without obvious physical limit.  

Of course, perfect neutron diodes do not grow on trees, and the challenge lies in creating a genuine unidirectionality of the neutron current so that the criticality $\keff$ does not increase (or only in a subleading way) when stacking more subcritical layers.

\begin{figure}
    \centering
    \includegraphics[width=0.99\columnwidth]{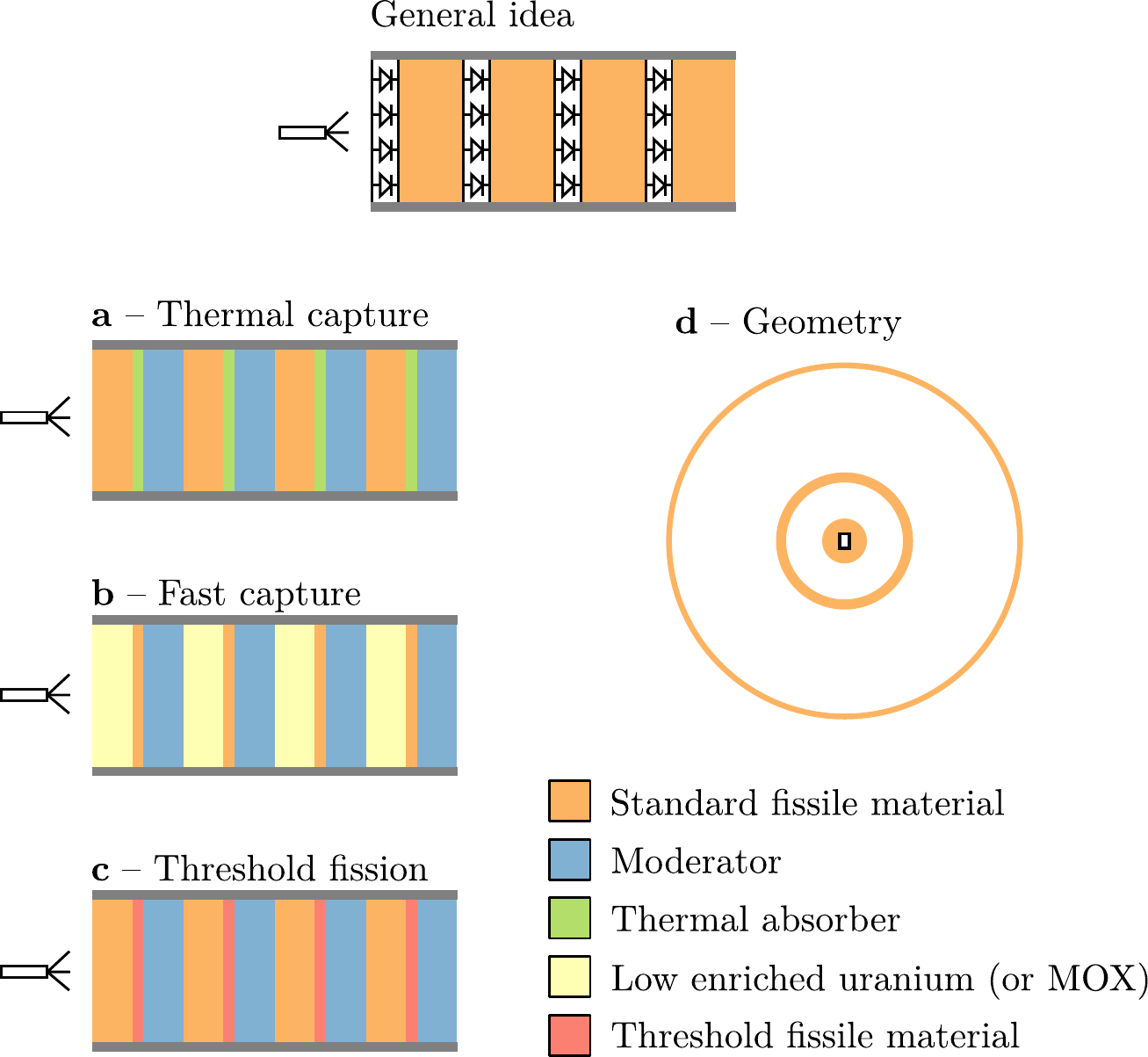}
    \caption{\textbf{Strategies for multi-stage amplification} -- Subcritical cells separated by ideal neutron diodes act as neutron amplifiers. In practice, this could be implemented with 4 distinct physical strategies.}
    \label{fig:strategies}
\end{figure}

There are at least four conceptually different ways to create a one way neutron current between different parts of a reactor, summarized in Fig. \ref{fig:strategies}. In practice, these different methods can be combined, and I separate them only to emphasize the physical mechanism at play.

The first idea corresponds to the one that, to my knowledge, was considered initially, as early as 1957 \cite{borst1957}. The idea is to exploit the fact that certain materials like cadmium are almost transparent to fast neutron but have a large capture cross section for thermal neutrons. Putting such a thermal absorber followed by a moderator, one obtains a material that can be crossed only one way by neutrons (and converts fast neutrons to slow neutrons in the forward direction).

The second option is in a way opposite to the first. Instead of capturing slow neutrons, one could selectively capture fast neutrons with a material like low enriched uranium that works as a neutron multiplier only in thermal spectrum while captures by \isotope[238]{U} dominate in fast spectrum. The idea is to put a moderator first and then this fast spectrum capture material to obtain a diode. This one way amplification through fast captures is likely insufficient to create a genuine irreversibility alone, but seems to play a role in the multiplication reached by various proposals.

The third option is to use a threshold fissile material like \isotope[237]{Np} or \isotope[241]{Am}, which is fissile in fast spectrum only and that consequently multiplies only fast neutrons. As before, this allows to construct a one way amplifier with the addition of a moderator.

The fourth option is purely geometric, and is in a way the simplest (see \cite{nifenecker2003accelerator}). The idea is to decouple stages by having them grow in size (or in surface) as one goes away from the source. This way, neutrons from an earlier stage can reach the following one where they start fissions, but these fissions create neutrons that have lower chances to hit the initial stage. An example consists in considering concentric fissile spheres of quickly growing radii and shrinking width. If the ratio of the radii is large enough, the total $\keff$ of the set of spheres is the same as that of each of them taken individually. However, the number of fissions is amplified at each layer.

Let us finally mention a more recent idea, which is to exploit magnetic fields to deflect neutrons via the coupling to their spin \cite{arias2016}. As it is far more speculative than the previous ones, I will not consider this option further here.

 \paragraph{Pre-amplifier exploration --}
 
 \noindent
My objective is to see how a combination of the previous strategies could fare with physically realistic materials and geometry (including neutron loss). To this end, it seems easier to distinguish two levels of amplification, that would be joined together in a complete reactor, but that come with starkly different constraints. A first stage, the preamplifier, would bring an arbitrarily faint neutron flux to the intensity needed to drive a second ``power'' stage. This second stage, a more standard reactor core, would be made of only a few simple layers, with low amplification but high power density. 

The proper study of a modestly amplifying reactor core is difficult but largely done in the literature \cite{seliverstov1996,kiselev2002}. The difficulty comes from the need to take into account thermohydraulic constraints (which limit the local power density), the evolution of reactivity, neutron economy, and simplicity constraints. Typical two layer cores exploit a threshold fissile material and concentric cylinders to achieve multiplication. For realistic designs, a modest amplification can be achieved, but is not sufficient to drastically change the neutron driving requirements.

To my knowledge, a many-layer preamplifier has never been studied. The constraints are different from those of the core. To achieve massive amplification, the design should be scalable, which \emph{a priori} excludes playing with a spherical or cylindrical concentric geometry, and favors a linear stacking of layers. On the other hand, the power density would be orders of magnitude lower than the one in the core, especially in the first layers closer to the (faint) source. This in turns means that cooling, reactivity evolution, and neutron economy would be secondary constraints. They should be taken into account ultimately, but in the spirit of a first exploration, it makes sense to neglect them to focus on neutron transport. Finally, such a preamplifier might be useful as a standalone device, \eg to produce intense neutron pulses (at low average power), or breed radioisotopes and tritium, which further motivates its study. 
\begin{figure}
    \centering
    \includegraphics[width=\columnwidth]{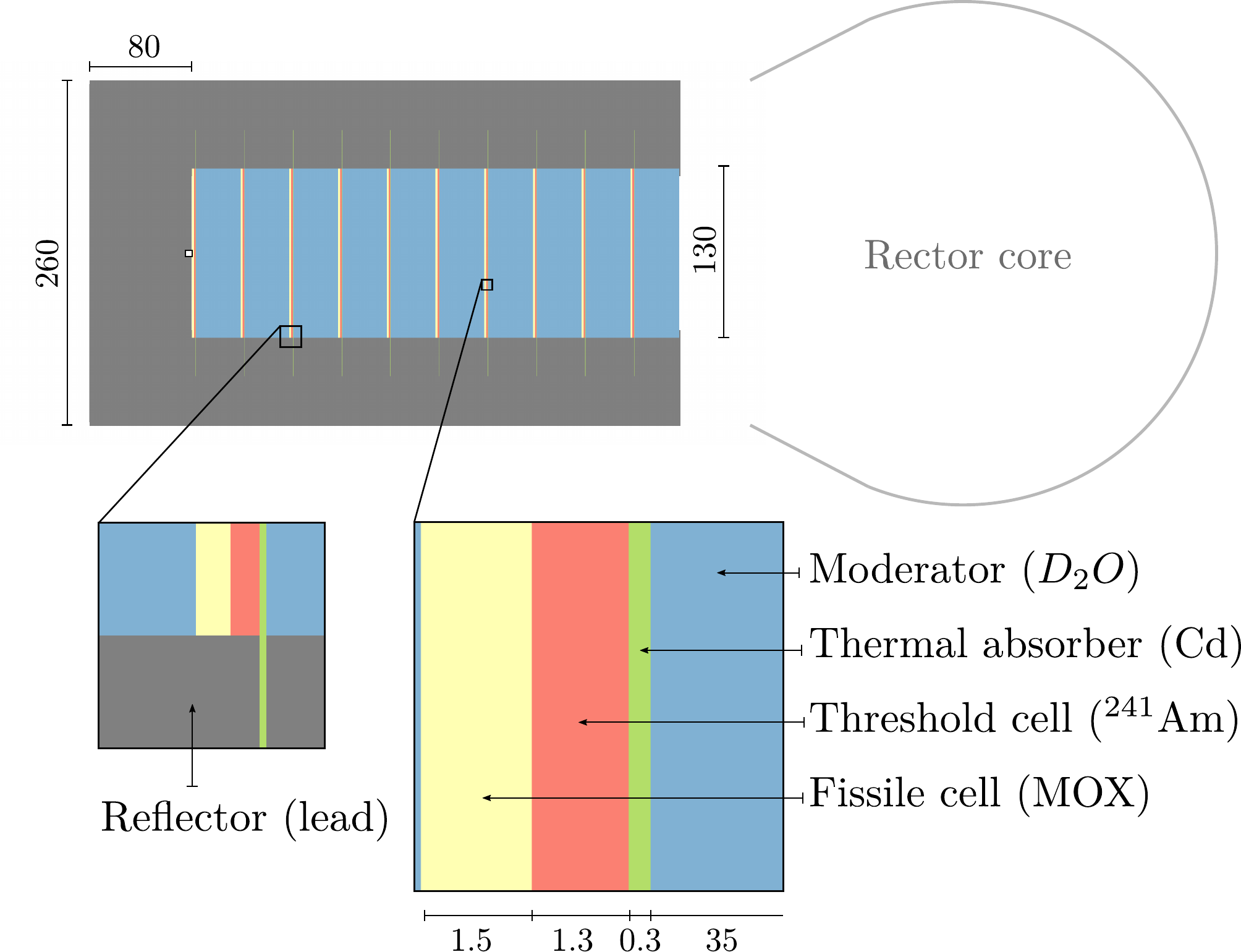}
    \caption{\textbf{Preamplifier design} -- The proposal is shown on a 10 layer example. The preamplifier is a cylinder with $130\times130~\text{cm}^2$ square basis surrounded by a lead reflector. The moderator is heavy water, the thermal absorber cadmium, the threshold material is metallic \isotope[241]{Am}, and the fissile cell MOX ($43\%$ \isotope[239]{Pu} and $57\%$ \isotope[238]{U} for $\keff=0.97$). The cadmium layer goes half-way through the reflector to reduce the backcoupling of thermal neutrons through the reflector. The neutron source is show by a white dot, against the reflector and just before the first layer. The dimensions are shown in $\mathrm{cm}$.}
    \label{fig:design}
\end{figure}

Finding an optimal or even adequate design of preamplifier given reasonable engineering and safety constraints is a non trivial task. My objective is more modest: I merely want demonstrate that amplification is physically feasible, at least in principle. To this end, I threw everything but the kitchen sink (geometry), to keep a scalable design where an arbitrary number of layers can be stacked. More precisely, I considered a stack of layers each made of 4 cells containing 1) a fissile material with fission enhanced in thermal spectrum [a mixed oxide of \isotope[239]{Pu} and \isotope[238]{U}], 2) a threshold fissile material [metallic \isotope[241]{Am}], 3) a thermal neutron absorber [cadmium], and 4) a moderator [heavy water]. This combines the 3 non geometric amplification methods mentioned before. This choice is largely arbitrary, and it is likely that much better combinations can be found. The precise geometry, including the reflector, is provided in Fig. \ref{fig:design} for a stack of 10 layers.
 
\paragraph{Results --}
 \noindent
 I carried a simple test of the previous proposal with the help of the Monte Carlo transport simulation tool openMC \cite{openmc} developed at MIT. 
 
\begin{figure}
     \centering
     \includegraphics[width=\columnwidth]{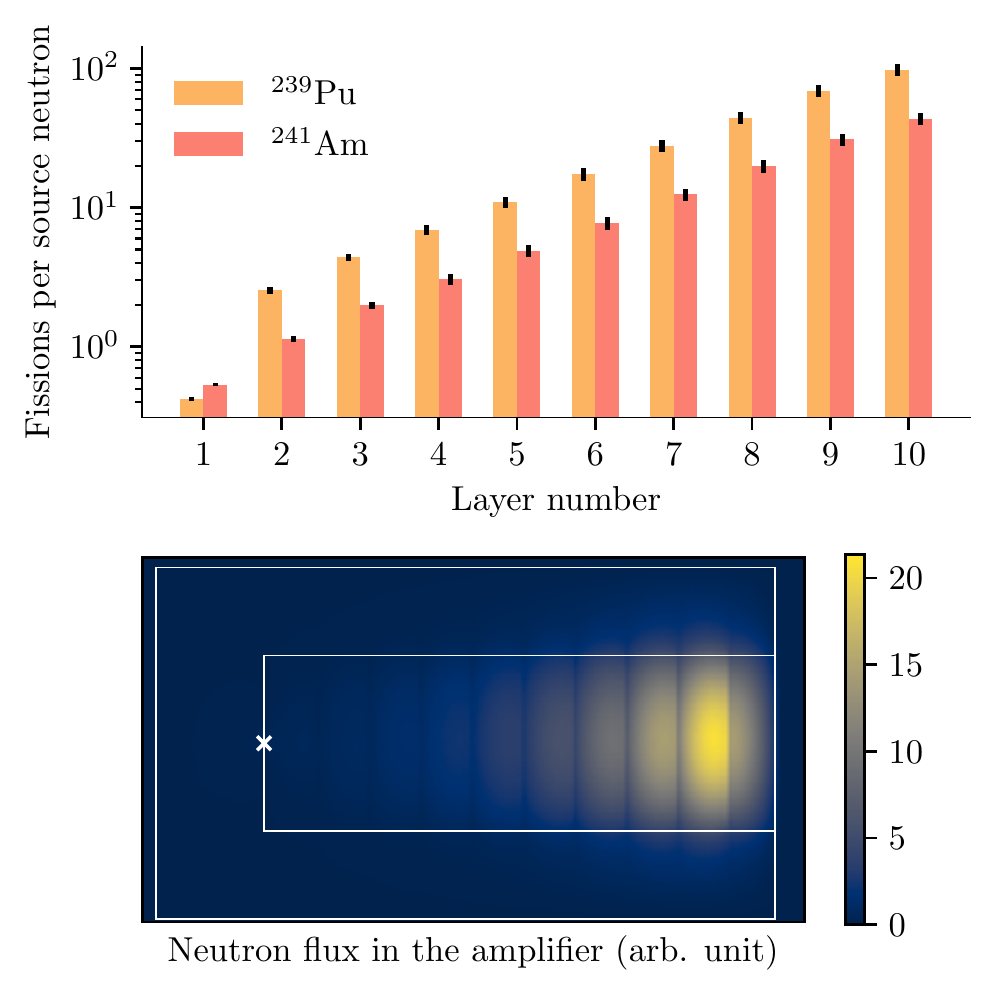}
     \caption{\textbf{Neutron amplification for a 10-layer system} -- After the first layer, the number of fissions per layer is approximately multiplied by $\times 1.6$ every layer. The neutron flux also increases geometrically as a function of the layer number, and the flux near the external source is consequently negligible compared to the flux at the end. The white cross shows the position of the source.}
     \label{fig:amplification}
 \end{figure}

 After some trial and error but no systematic optimization, it is easy to reach a scalable amplification of $ \simeq 1.6$ per layer [1.6 times more neutron flux and fissions in layer $n+1$ than in layer $n$] for $\keff=0.97$ while keeping layers with a reasonable cross section $130\times 130~\text{cm}^2$ and width $\simeq 38~\text{cm}$ (see Fig. \ref{fig:amplification}). Let us briefly discuss the trade offs involved. 
 
 Most of the space is taken by the heavy water moderator, with a thickness of $35~\text{cm}$ per layer. The thicker this cell, the better the neutron flux is thermalized which enhances the decoupling between layers. On the other hand, a thicker cell increases the leak of thermal neutrons into the lead reflector where they can recouple to previous layers. Taking a larger layer cross section allows to proportionally reduce the surface leaks in the reflector and thus permits an increase in the moderator thickness. This yields an improved amplification at fixed $\keff$, at the price of an overall larger amplifier. Using light water would reduce the thickness required to thermalize the neutron flux, but I observed a lower overall amplification because of the higher capture rate.
 
 The mixed oxide section is chosen rather narrow ($1.5~\text{cm}$) and highly fissile\footnote{I used isotopically pure \isotope[239]{Pu} for simplicity, as it provides the highest amplification. But there is no objection to the use of reactor grade plutonium, which would allow amplification as well.} ($43\%$ \isotope[239]{Pu}). This only weakly exploits the second strategy of fast capture by \isotope[238]{U}. This is because the backward decoupling effect brought by \isotope[238]{U} is out-weight by the increased coupling between layers allowed by a thinner fissile cell. It is still possible to reach some amplification (albeit weaker) by opting for a thicker section with lower \isotope[239]{Pu} content (\eg $8\%$). This then has the advantage of keeping the system subcritical even in the event of a complete emptying of the moderator (whereas for higher \isotope[239]{Pu} contents, the reactivity first decreases after a partial loss of moderator, but reincreases after a full emptying, as layers recouple backward).
 
 The americium section is taken similar in thickness to the fissile one ($1.3~\text{cm})$ and allows to fully exploit the third strategy. Finally, the cadmium layer is very thin ($3~\text{mm}$) which is sufficient to capture most thermal neutrons while remaining mostly transparent to epithermal and fast ones. This implements the first strategy. 

\begin{figure}
    \centering
    \includegraphics[width=\columnwidth]{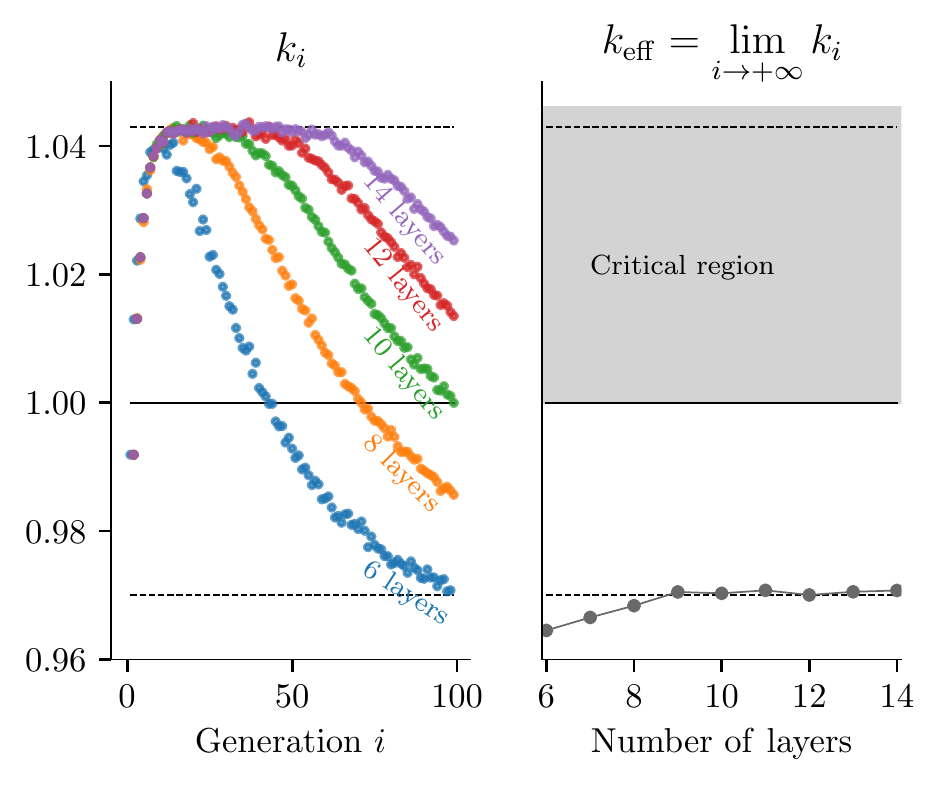}
    \caption{\textbf{Transient neutron multiplication} -- The number of neutrons produced per generation $k_i$ reaches a plateau at $k\simeq 1.04$, that lasts longer for a larger number of layers. The system remains subcritical no matter the number of layers, with $\keff\rightarrow 0.97$ up to measurement errors.}
    \label{fig:k}
\end{figure}
 
The way arbitrary geometric amplification is allowed despite fixed $\keff$ is transparent in this proposal and follows the argument in the introduction. Stacking more layers  does not bring the system closer to criticality (after a certain point) as backward coupling is suppressed (see Fig. \ref{fig:k}). However the neutron multiplication $k_i$ remains larger than $1$ for a number of generations proportional to the number of layers, and passes below $1$ only once neutrons start leaking at the end of the amplifier. For 10 layers, this happens only after about 100 generations.
 
 \paragraph{Discussion --}
 
 \noindent
 With the rudimentary amplifier I put forward, it is possible to reach an amplification as large as one needs. Since the amplification grows geometrically in the number of layers, one can in principle have as input an arbitrarily faint neutron source, and even dream of ditching the accelerator driven spallation source to use radioisotopes. 
 
Naturally, this comes with severe limitations that are yet to be addressed. Assuming a realistic amplifier could be designed along the lines presented here (including cladding, cooling, control systems, etc.), its safety would still need to be demonstrated. Indeed, the preamplifier operates away from its maximally reactive geometry: a sufficient bending of the layers or melting of the cells could turn the system critical. Further, one would need an efficient \emph{in situ} measurement of $\keff$: in all ADSRs but even more so in a multi-layer system, the output power can decrease as $\keff$ increases, making subcriticality difficult to guarantee in practice. The development of accurate reactivity measurement techniques is an active area of research (see \eg \cite{chevret2014,marie2019}). 

A more minor drawback of a subcritical system with a large preamplification, compared to standard ADSRs, is that the neutron economy is unavoidably worse, as input neutrons are negligible in the overall balance. This means, at least \textit{a priori}, that thorium breeding in thermal spectrum is more difficult with such designs.

I have studied the preamplifier as a standalone device, and its use to drive a reactor core would necessarily bring some residual backward neutron coupling, slightly increasing the $\keff$ of the whole. The quick saturation of the $\keff$ as the number of preamplifying layers is increased  (see Fig. \ref{fig:k}) suggests that applying the same decoupling strategy and seeing the core merely as a last, much larger layer, should be sufficient to ensure global subcriticality. This should however be demonstrated in full fledged neutron transport simulations once a more complete and realistic design is proposed. 

These limitations notwithstanding, there are interesting open questions of a more theoretical nature. In particular, it would be helpful to know what the maximum reachable amplification is for given $\keff$, reactor size budget, and safety constraints. Likewise, what is the lowest $\keff$ that can be reached with scalable amplification? Multi-layer systems also present challenges for accurate Monte-Carlo simulations: naively estimating $\keff$ and the neutron flux everywhere in the system becomes prohibitively costly as the number of layers is increased, because of the transient exponential gain. There is certainly a way to do better.

In the end, scalable subcritical amplification presents great challenges but is physically sound. It is in principle feasible with materials that exist, and with a simple geometry. If realized, such amplifiers could become a useful building block of full fledged reactors, to produce electricity, incinerate (some) radioactive waste, or provide on demand sources for science. As a result, multi-layer amplification likely deserves to be studied further and at a finer level of detail. I hope the present paper will stimulate this exploration.

\begin{acknowledgments}
I am grateful to Sylvain David and Alexis Nuttin for helpful comments and discussions.
\end{acknowledgments}

\bibliographystyle{apsrev4-1}
\bibliography{main}
\end{document}